\documentclass[twocolumn,preprintnumbers,superscriptaddress,prl,nofootinbib,longbibliography,amsmath,amssymb]{revtex4-1}
\usepackage{graphicx}% Include figure files
\usepackage{times}
\usepackage{color}
\usepackage[version=4]{mhchem}
\definecolor{darkblue}{rgb}{0,0.02,0.45}
\usepackage[colorlinks=true,citecolor=darkblue]{hyperref}
\usepackage{dcolumn}% Align table columns on decimal point
\usepackage{bm}% bold math
\usepackage{ulem}
\usepackage{gensymb}
\usepackage{braket}
\usepackage{amsmath}
\usepackage[percent]{overpic}

\definecolor{zm}{rgb}{0,0.5,1}
\usepackage{multirow}
\usepackage{xfrac}

\begin{document}

%This manuscript has been authored by UT-Battelle, LLC, under Contract No. DE-AC0500OR22725 with the U.S. Department of Energy. The United States Government retains and the publisher, by accepting the article for publication, acknowledges that the United States Government retains a non-exclusive, paid-up, irrevocable, world-wide license to publish or reproduce the published form of this manuscript, or allow others to do so, for the United States Government purposes. The Department of Energy will provide public access to these results of federally sponsored research in accordance with the DOE Public Access Plan (http://energy.gov/downloads/doe-public-access-plan).
%\clearpage

\title{Structure transition and zigzag magnetic order in Ir/Rh-substituted
honeycomb lattice $\alpha$-RuCl$_3$}

\author{Zachary Morgan}
\affiliation{Neutron Scattering Division, Oak Ridge National Laboratory,  Oak Ridge, Tennessee 37831, USA}

\author{Iris Ye}
\affiliation{Next Generation STEM Internship Program Participant}

\author{Colin L. Sarkis}
\affiliation{Neutron Scattering Division, Oak Ridge National Laboratory,  Oak Ridge, Tennessee 37831, USA}

\author{Xiaoping Wang}
\affiliation{Neutron Scattering Division, Oak Ridge National Laboratory,  Oak Ridge, Tennessee 37831, USA}

\author{Stephen Nagler}
\affiliation{Neutron Scattering Division, Oak Ridge National Laboratory,  Oak Ridge, Tennessee 37831, USA}
\affiliation{Department of Physics and Astronomy, University of Tennessee, Knoxville, Tennessee 37996, USA}

\author{Jiaqiang Yan}
\affiliation{Materials Science and Technology Division, Oak Ridge National Laboratory,
Oak Ridge, Tennessee 37831, USA}
\date{\today}

\begin{abstract}
We report magnetization and neutron diffraction studies on crystal and magnetic structures of
Ir- and Rh-substituted honeycomb lattice $\alpha$-RuCl$_3$. The iridium or rhodium atoms are
distributed at the Ru site with little structural modification. Both systems
undergo a room-temperature monoclinic $C2/m$ to low-temperature trigonal $R\bar{3}$
phase transformation with a large recoverable hysteresis. At low temperature,
a zigzag spin order is observed with the same characteristic wavevector $(0,0.5,1)$ as
in the parent $\alpha$-RuCl$_3$. Detailed magnetic structure refinement
reveals an ordered moment of $\rm 0.32(5) \mu_B/Ru$ and a upper boundary of canting angle
of $15(4)^\circ$ away from the basal plane at 5~K for the 10\% Ir-substituted $\alpha$-RuCl$_3$,
which is different from the 0.45-0.73~$\rm \mu_B/Ru$ and $32^\circ$-$48^\circ$ canting angle
reported in the parent compound $\alpha$-RuCl$_3$. The observation of unchanged RuCl$_6$
local octahedral environment, reduced magnetic moment size and canting
angle highlights the potential to study quantum spin liquid behavior through non-magnetic ion doping.

\end{abstract}

\maketitle

\section{Introduction}
The interplay between topology and electron correlation leads to a large variety of
interesting topological phenomena in quantum materials. These include fractional quantum
Hall effect where the quasi-particles carry a fraction of the electron charge \cite{du09,bolotin09} and
quantum spin liquid (QSL) where the quantum fluctuations prevent magnetic order down
to absolute zero temperature \cite{balents10,kitagawa18}. The Kitaev spin model on a
honeycomb lattice \cite{kitaev06} is an exactly solvable example in which spin dynamics fractionalizes
into itinerant Majorana fermions and Ising gauge-field excitations.
The experimental realization for this model was initially proposed in quasi-2D iridates
$\rm A_2IrO_3$ \cite{jackeli2009}, where the bond directional Ising interactions arises
from spin-orbit-assisted $j_{\rm eff}=1/2$ state in an edge-sharing environment.
Recent focus has gradually shifted to the honeycomb lattice
$\alpha$-RuCl$_3$ \cite{plumb14, takagi2019concept}.
Despite the apparent appearance of long-range magnetic order,
both inelastic neutron scattering and Raman spectroscopy
\cite{sandilands15,nasu16,banerjee16,banerjee18}
have detected a broad scattering continuum in $\alpha$-RuCl$_3$.
This continuum aligns with the predicted presence of
itinerant Majorana fermion and is interpreted as indicating that $\alpha$-RuCl$_3$ is situated in close proximity to the QSL state. Additionally,
anomalous thermal Hall properties are observed in the QSL state with applied in-plane magnetic fields \cite{kasahara18,yokoi21,bruin22,czajka23}.

Below $T_N=7$~K,  $\alpha$-RuCl$_3$ exhibits a
zigzag spin structure \cite{park16,cao16,balz21} like what was observed in honeycomb
iridates \cite{choi12,ye12a,chaloupka13}. The static spin configuration is quite fragile and susceptible to
various perturbations including in-plane magnetic field \cite{sears17,baek17,banerjee18}, hydrostatic
pressure \cite{wolf22,wang23,bhattacharyya23}, stacking disorder \cite{johnson15,cao16},
and chemical substitution \cite{lampen-kelley17,bastien19,bastien22}.
In the latter case, partial substitution of Ru$^{3+}$ with magnetic Cr$^{3+}$ ions
destabilizes the zig-zag long-range magnetic order and favors
a spin-glass state \cite{bastien19}. Alternatively, the inclusion of nonmagnetic ions such
as Rh$^{3+}$ \cite{bastien22} and Ir$^{3+}$ \cite{lampen-kelley17} seems to only suppress
the zigzag order without any signature of freezing into a spin glass state.
Furthermore, inelastic neutron scattering studies have found that the key character of fractionalized
excitations is maintained in a wide range of Ir concentration even when the long-range
spin order is absent \cite{lampen-kelley17}. This is particularly intriguing as
it reveals the robust nature in the dynamical channel and provides a viable route to
achieve the long-sought-after quantum spin liquid state in the absence of applied
magnetic field.

The impact of chemical substitution on the crystal and magnetic
structures of $\alpha$-RuCl$_3$ has not been thoroughly studied before partially
due to the availability of sizable single crystals. In this work, the growth, magnetic
properties, nuclear and magnetic structures of Ru$_{1-x}{R}_x$Cl$_3$ ($\rm {\it R}=Ir, Rh$)
single crystals are reported. With a moderate concentration of $x=0.1$, the transition temperature $T_N$ is suppressed to
around 6~K. A large hysteresis about 70~K is present between the room temperature ($T$)
monoclinic $C2/m$ and the low-$T$ trigonal $R\bar{3}$ crystal structures
likely due to the presence of stacking disorder. As observed in $\alpha$-RuCl$_3$,
both systems revert to the room temperature monoclinic structure after thermal cycling without sign of stacking disorder
evident as diffuse scattering along the $c^\ast$-axis. A significant reduced moment of 0.32(5)~$\mu_B/\rm Ru$ is
observed in the Ir-substituted $\alpha$-RuCl$_3$ zigzag spin configuration. The spin moment is
in the $ac$-plane with tilting angle 15(4)$^\circ$ away from the $a$-axis [the direction perpendicular to the Ru-Ru bond].
Both moment size and canting angle are smaller than previous reports in $\alpha$-RuCl$_3$
with moment size of 0.45-0.73~$\mu_B/\rm Ru$ and canting angle of 35-48$^\circ$ \cite{cao16,park16}.

\begin{figure}[!ht]
   \includegraphics[width=3.4in]{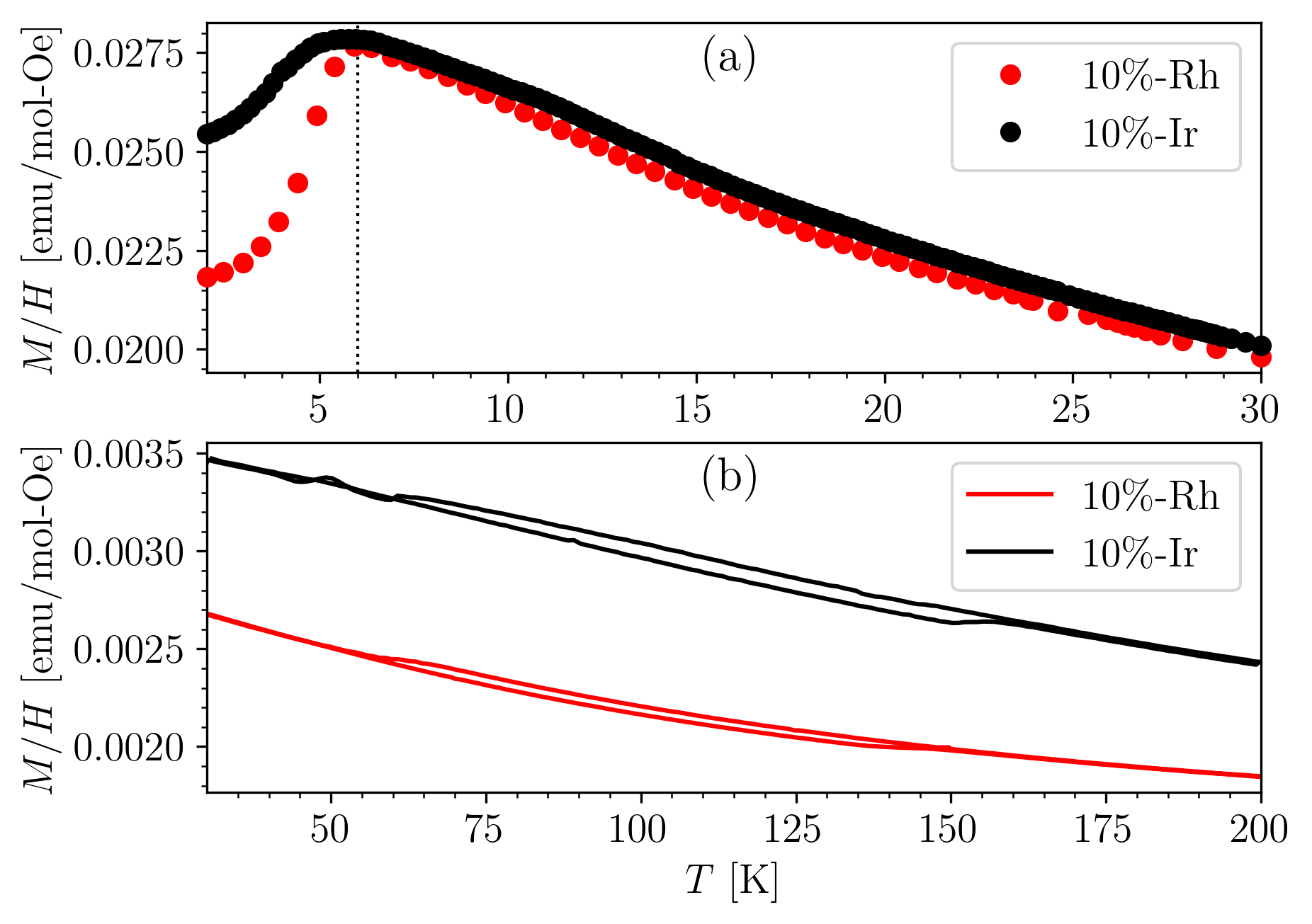}
   \caption{Temperature dependence of magnetization of 10\% Ir- and Ru-substituted $\alpha$-RuCl$_3$. (a)
   Low temperature magnetization highlighting the magnetic ordering temperature. The data
   were collected in a magnetic field of 1~kOe along an arbitrary direction in $ab$-plane.
   (b) High temperature magnetization highlighting the structural transition at high temperatures.
   The measurement is performed with a ramping rate of 2\,K/min in a magnetic field of 10~kOe applied perpendicular to the plate.
   }
\label{Mag}
\end{figure}

\section{Experimental Details}

Single crystals of Rh- and Ir-substituted $\alpha$-RuCl$_3$ were prepared using self-selecting vapor
growth method \cite{yan23}. The commercial RuCl$_3$ powder from Furuya Metals (Japan) was
used in the growths without further purification. Pure RhCl$_3$ and IrCl$_3$ powders were obtained by
reacting the transition metal oxides with AlCl$_3$-KCl salt \cite{yan17}. Plate-like
crystals with a typical in-plane dimension of 5-10~mm and thickness up to 1~mm can be
obtained. Magnetic properties were measured with a Quantum Design (QD) Magnetic Property
Measurement System in the temperature range 2.0-300~K. The compositions of the crystals
were determined using energy dispersive spectroscopy (EDS) with a Hitachi TM3000 scanning
electron microscope and Bruker Quantax70 spectrometer. EDS suggests the transition
metal/chlorine ratio is 1:3. However, the significant overlap of peaks prevents a reliable
determination of the real composition. Therefore, nominal composition
is used in this work.  Magnetic measurements were used to screen crystals
for further characterizations. We noticed some variation of magnetic and
structural transitions among crystals from the same batch, indicating the
variation of substitution content. Therefore, magnetic and neutron
diffraction measurements were performed on the same piece of
crystals for each composition.

Neutron diffraction experiments of 10\% substituted compositions were carried out using
the single crystal diffractometer CORELLI \cite{ye18} at the Spallation Neutron Source (SNS) to study
both the crystal and magnetic structures and the corresponding structural
transition. Individual crystals were glued to a thin
aluminum plate and loaded inside a closed-cycle refrigerator with a base temperature of 5~K.
The neutron diffraction data are first collected at 200~K (above the structural transition) with 360 degrees of rotation about the vertical axis.
After samples were cooled to base temperature, another full map
was collected to investigate the magnetic structure. The crystals were then warmed
back to 200~K to probe possible structural change after thermal cycling. During initial cooling and
subsequent warming, characteristic Bragg peaks that are unique to the specific space group of the crystal were
closely monitored as function of temperature with a ramping rate of 1~K/min.

A similar single crystal neutron diffraction experiment
of 20\% Rh-substituted samples were also performed
on the TOPAZ diffractometer \cite{schultz2014integration,coates2018suite} at SNS.
Crystals were attached via GE varnish
to thin aluminum posts with the monoclinic $a^*$ axis oriented vertically. Sample temperature was
controlled by a Cryomech P415 pulse tube cryocooler and data were collected using sample orientations
optimized for coverage with the CrystalPlan software \cite{zikovsky2011crystalplan}.
Data were taken above the structural transition at 250~K as well as below the structural transitions
at base temperature of 15~K. Crystals were cooled and heated at a rate of 2~K/min and the
transitions were tracked studying the nominal $(1,1,3)$ Bragg peak of the low temperature $R\bar{3}$ structure.

\begin{figure}[!ht]
   \includegraphics[width=3.4in]{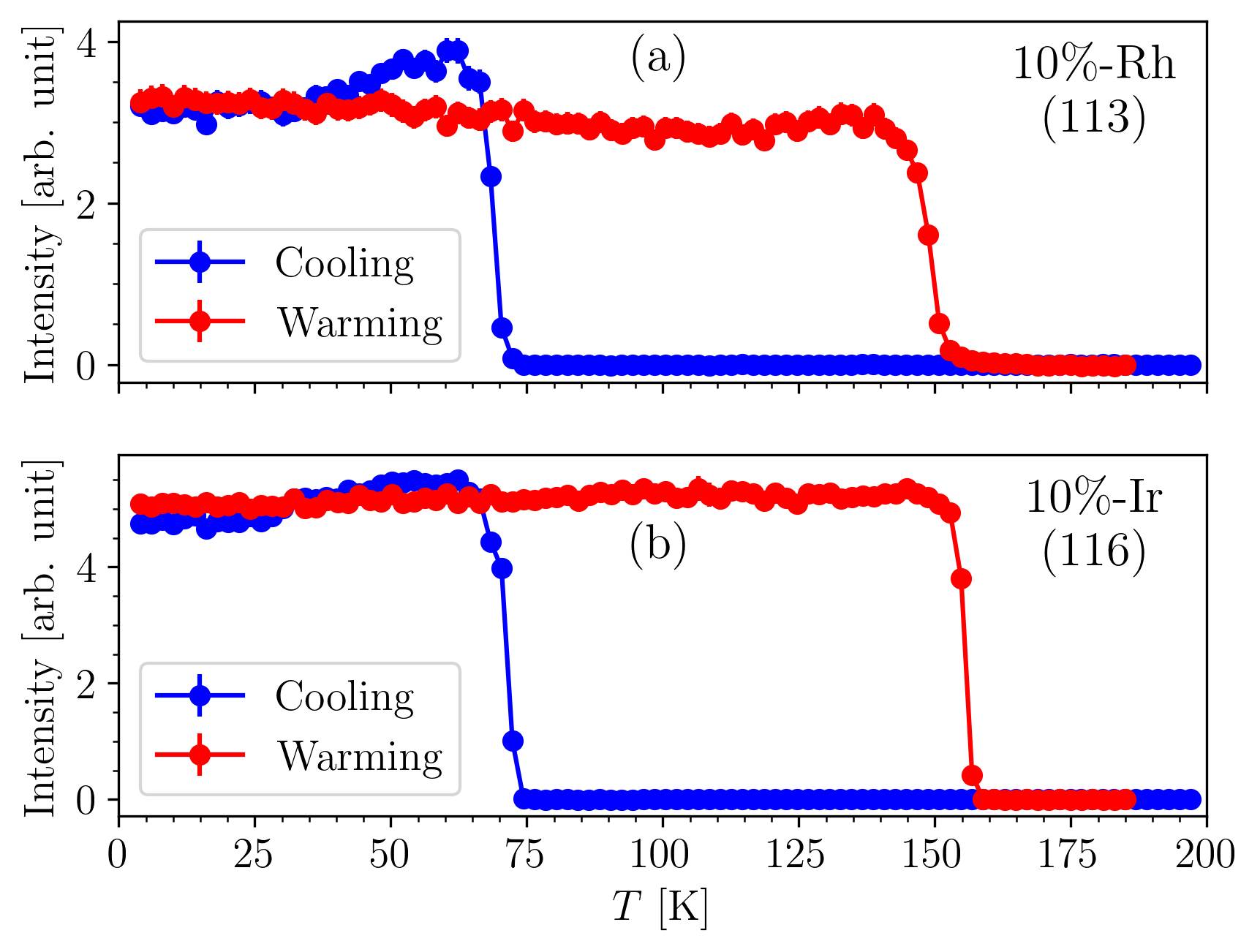}
   \caption{Cooling (blue) and warming (red) curves of integrated peak intensity for (a) 10\% Rh-substituted and
   (b) 10\% Ir-substituted $\alpha$-RuCl$_3$ showing large hysteresis in structure transformation between high-temperature
   $C2/m$ monoclinic and low-temperature $R\bar{3}$ trigonal structure.
   The first order transition is complete and does not exhibit remnant phase coexistence at low
   temperature.}
\label{hysteresis-1}
\end{figure}

\begin{figure}[!ht]
   \includegraphics[width=3.4in]{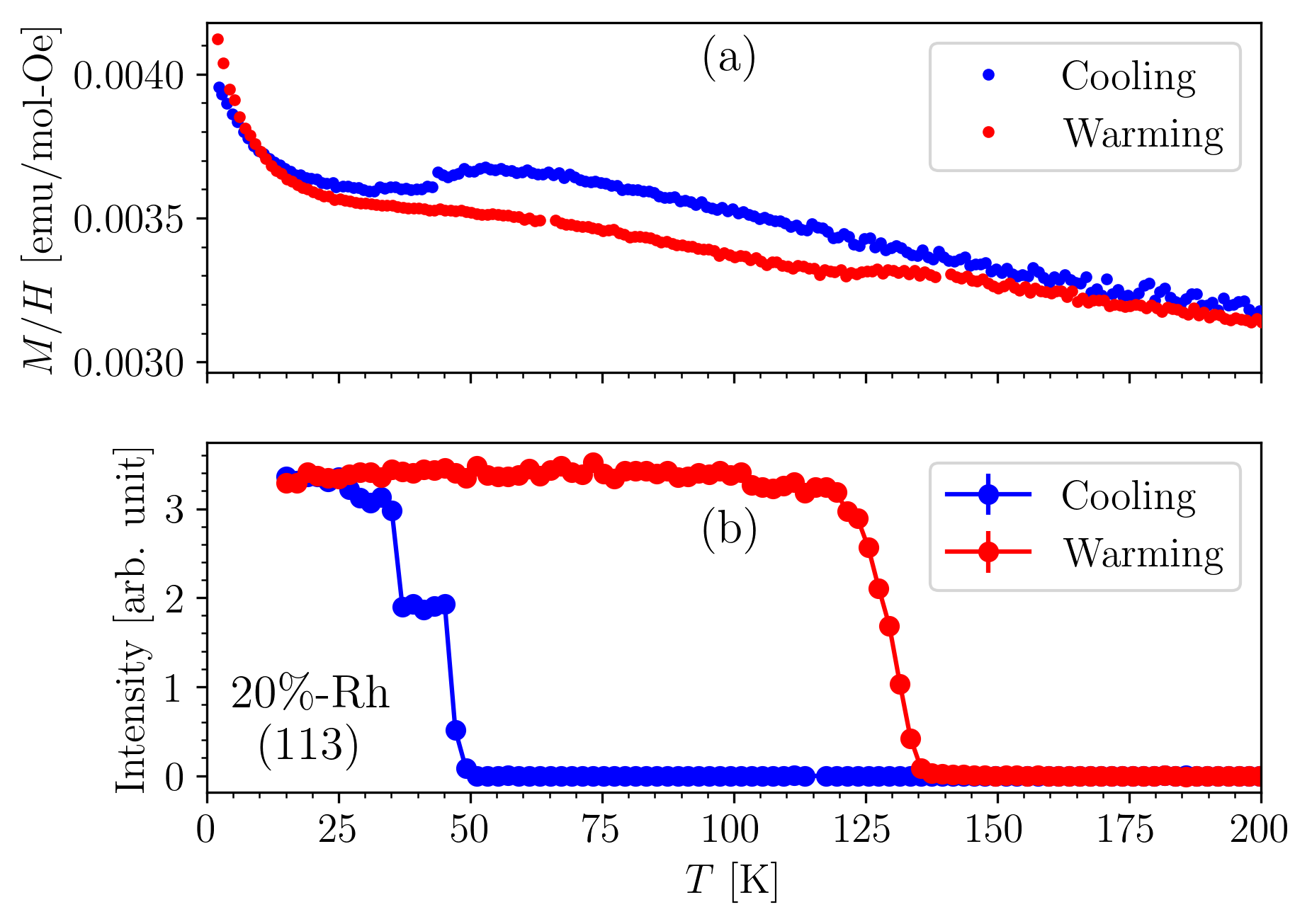}
   \caption{Temperature dependence of (a) magnetization and (b) intensity of $(1,1,3)$ reflection of 20\% Rh-substituted $\alpha$-RuCl$_3$. Magnetic data were collected with a ramping rate of 2\,K/min in a magnetic field of 1~T applied perpendicular the crystal plate. No long range magnetic order is observed above 2\,K while the structure transition is still present.
   }
\label{Rh20}
\end{figure}

\section{Results and Discussion}

\begin{figure*}[!ht]
   \includegraphics[width=6.8in]{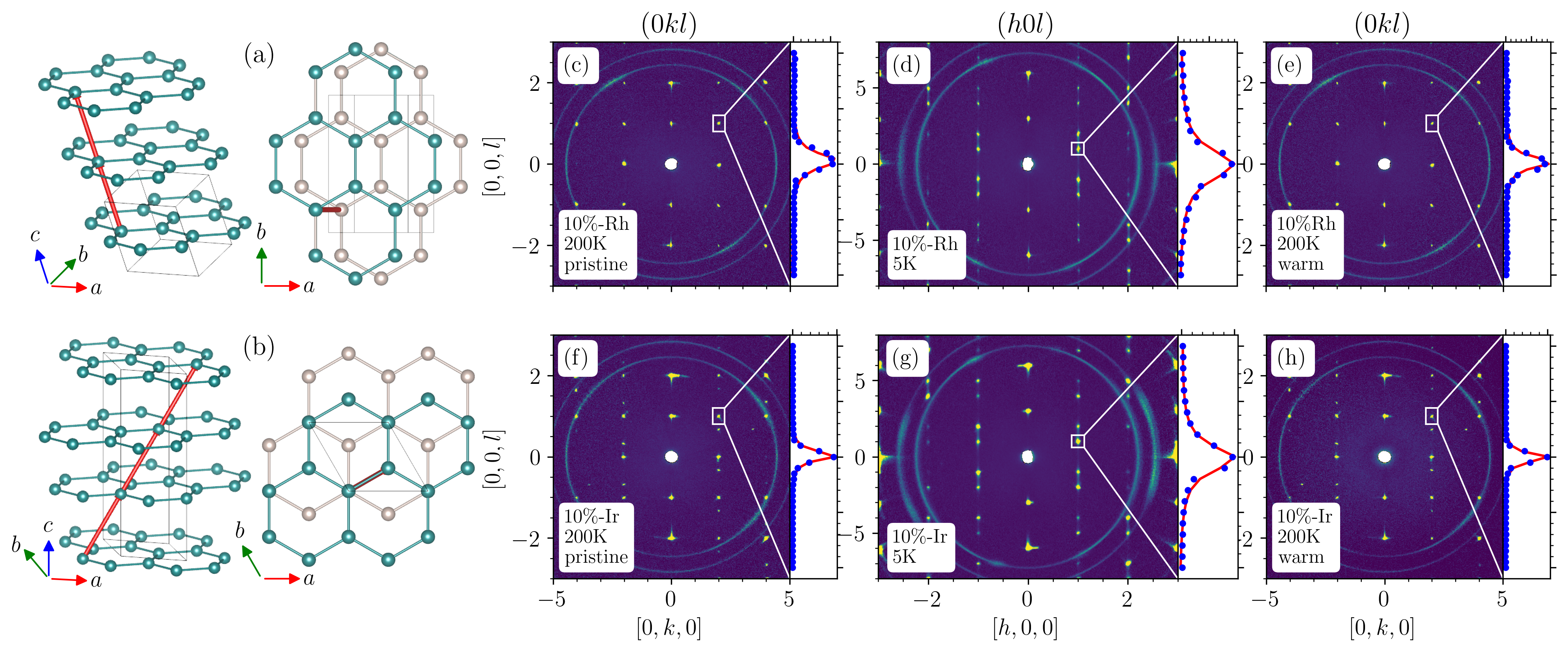}
   \caption{Stacking sequence of honeycomb layers in (a) $C2/m$ and (b) $R\bar{3}$ space groups.
   The out-of-plane stacking shift vector is $[0,0,1]$ and $[\sfrac{2}{3},\sfrac{1}{3}, \sfrac{1}{3}]$,
   respectively. Heatmap plot of 10\% Rh-substituted $\alpha$-RuCl$_3$ at (c) 200~K on first cooling, (d) 5~K,
   and (e) 200~K after warming. Similar data collected for 10\% Ir-substituted $\alpha$-RuCl$_3$ at (f) 200~K on
   cooling, (g) 5~K, and (h) 200~K on warming.
   Both Rh- and Ir-substituted crystals are
   indexed using monoclinic $C2/m$ space group at high temperature while the low-$T$ data are indexed in the hexagonal
   $R\bar{3}$ space group. Note both systems recover to the pristine state showing absence of
   structural diffuse scattering after thermal cycling back to 200~K. For the Ir-substituted sample, the
   appearance of an additional domain can be indexed using a twin with 120$^\circ$
   rotation about the $c^\ast$-axis. This domain recovers after thermal cycling. The inset shows selected line-cuts along the $c^\ast$-axis
   with sharp Bragg peaks in the monoclinic phase and broad peaks with diffuse scattering
   along $c^\ast$ in the trigonal setting.}
   \label{intensity-maps}
\end{figure*}

The temperature-dependent magnetization of 10\% Ru-
and Ir-substituted $\alpha$-RuCl$_3$ crystals used for neutron diffraction study at CORELLI is shown in Fig.~\ref{Mag}.
Both samples exhibit a long-range magnetic order with
similar $T_N=6$~K. The different magnitudes of the magnetization drop
below $T_N$ may result from different in-plane field directions. A weak hump around 10~K is observed for
the Ir-substituted crystals, indicating the presence of stacking disorder in this piece of
crystal. The higher temperature magnetization shown in \ref{Mag}(b) highlights the structure transition.
For both compositions, the transition occurs around 60~K
upon cooling and 150~K upon warming, resulting in a wide loop in the temperature
dependent magnetization.

The presence of the structure transition is further confirmed by neutron diffraction study.
Structural refinement indicates that the substituted $\alpha$-RuCl$_3$ systems crystallize
in a monoclinic $C2/m$ space group at room temperature
and transforms to trigonal $R\bar{3}$ space group at low temperature
consistent with other reports in crystals without doping \cite{mu22,zhang23}.
Figure~\ref{hysteresis-1}(a) shows the $T$-dependence of the nuclear reflection $(1,1,3)$
indexed in the low-$T$ $R\bar{3}$ space group upon cooling and warming.
The abrupt increase in intensity near 75~K of this peak on cooling and sudden disappearance at 150~K
on warming confirms the transition is first order and is consistent with previous investigation on the
pure $\alpha$-RuCl$_3$ where the temperature difference
of the structural transition between cooling and warming critically depends on the
degree of stacking disorder \cite{zhang23}. A narrower hysteresis window typically
implies the single crystal contains smaller amount of stacking disorder. The large
hysteresis windows observed in Figs.~\ref{Mag} and \ref{hysteresis-1} might reflect increased stacking
disorder. It is yet to be investigated whether the stacking disorder
comes from inappropriate
growth parameters or chemical substitution.

In parent $\alpha$-RuCl$_3$, the structure transition occurs around 170~K upon warming.
In contrast, our observations in this study indicate a transition temperature roughly 20~K lower
for 10\% Rh or Ir substituted compositions. This reduction in the structure transition temperature
appears to be a consequence of chemical substitution. Figure~\ref{Rh20}(a) shows the temperature
dependence of magnetization of 20\% Rh-substituted RuCl$_3$. No long range magnetic order was observed above
2~K for this composition. The thermal evolution of the $(1,1,3)$ peaks shown in Fig.~\ref{Rh20}(b)
indicates the structure transition occurs around 50~K upon cooling and 130~K upon warming.
Both temperatures are consistent with those obtained from magnetic measurements.
It is worth noting that the transition temperature upon warming is further reduced compared to those for 10\% Rh or Ir-substituted compositions.
This indicates that the nonmagnetic substitution in the honeycomb plane affects
not only the in-plane magnetic interactions but the inter-layer coupling.
Despite the apparent difference in the thermal hysteresis, the structural parameters among all
substituted $\alpha$-RuCl$_3$ samples from structure refinement remains unchanged compared to the
parent compound. The lattice constants, bond distances, and bond angles are nearly the
same within uncertainty (see Appendix).
This could be expected as the effective ionic radii of and Rh$^{3+}$ is 0.665~$\rm \AA$,
similar to Ir$^{3+}$ and Ru$^{3+}$ that are identical at 0.68~$\rm \AA$ \cite{shannon76}.

The difference between the room temperature monoclinic $C2/m$ phase shown in Fig.~\ref{intensity-maps}(a) and
low-$T$ trigonal $R\bar{3}$ phase shown in Fig.~\ref{intensity-maps}(b)
can be visualized in the reciprocal space mappings that are displayed Fig.~\ref{intensity-maps}(c)-Fig.~\ref{intensity-maps}(h).
Lattice parameter determination from observed structural
Bragg peaks indicates room temperature monoclinic cell parameters
with base centering conditions. At low temperature, a hexagonal cell is
observed with peaks indexed with rhombohedral lattice centering.
Each space group can be confirmed through
detailed analysis of the Bragg intensities using least squares refinement.
More details about the structure refinements and Bragg integration are given in the
Appendix.
In the room temperature phase, the honeycomb layers are stacked in an arrangement corresponding
to a simple translation along the $c$-axis with displacement vector perpendicular to the armchair of
the honeycomb bond that is illustrated in Fig.~\ref{intensity-maps}(a).
In the case of the low-$T$ trigonal phase, the stacking sequence
is different as shown in Fig.~\ref{intensity-maps}(b).
Viewing from the direction perpendicular to the basal plane, the neighboring layers are separated by an
in-plane displacement vector of $[2/3a, 1/3a]$ which is 60$^\circ$ from the armchair bond.

\begin{figure}[!th]
   \includegraphics[width=3.4in]{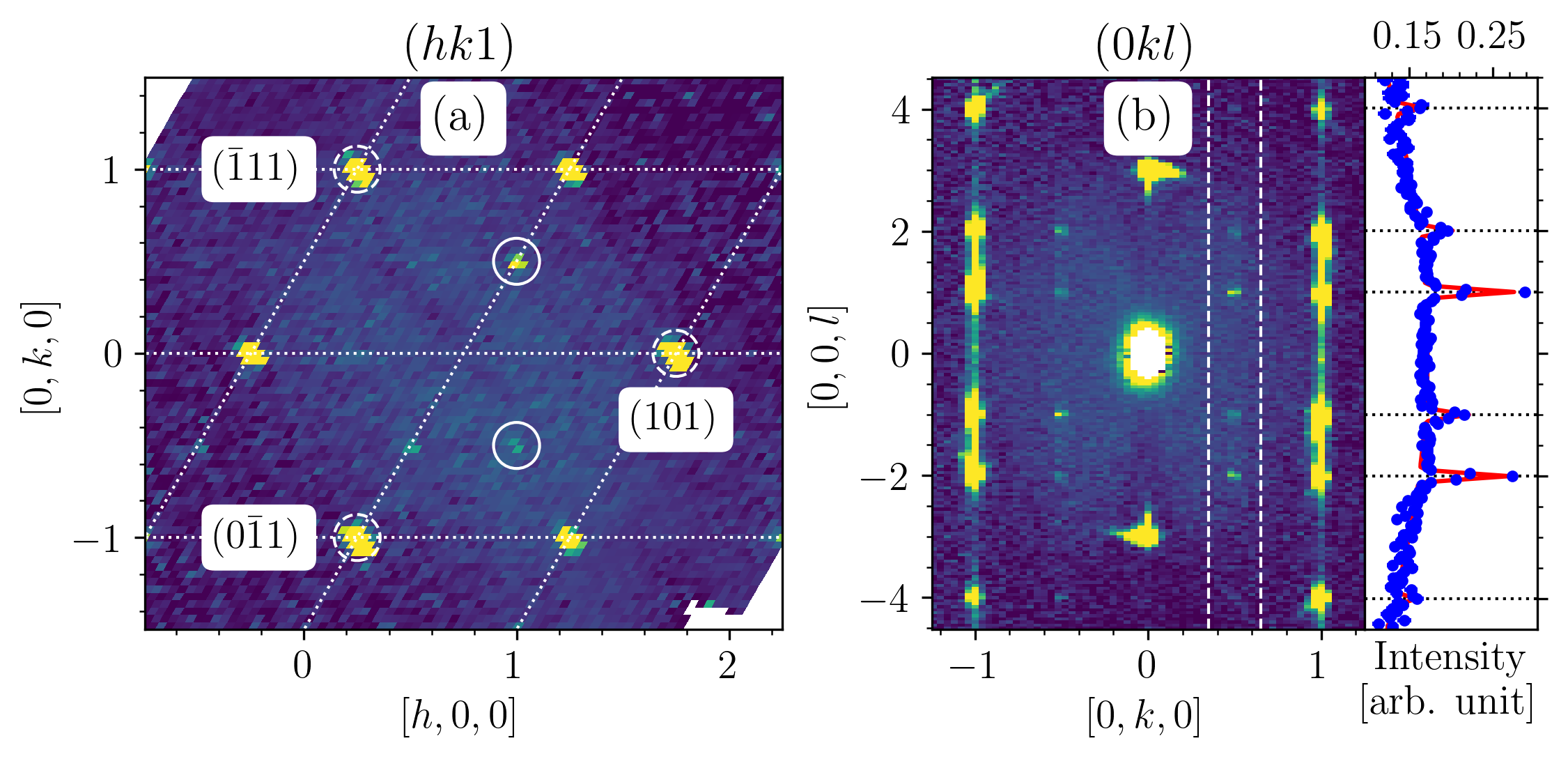}
   \caption{(a) Intensity slice in the $(h,k,1)$ plane showing the magnetic satellite peaks with
   propagation wavevector $(0,0.5,1)$ for the 10\% Ir-substituted $\alpha$-RuCl$_3$ at $T=5.0$~K.
  Two structural domains correspond to obverse/reverse twinning with volume ratio of 77:23.
  Structural peaks from the major domain are highlighted in dashed white circles.
  The magnetic twins associated with the major structural domain are evident with 120$^\circ$ rotation
  along $\boldsymbol{c}^\ast$ and marked in closed white circles.
  Note that the axes are skewed according to the 60$^\circ$ angle between
  $\boldsymbol{a}^\ast$ and $\boldsymbol{b}^\ast$.
  (b) The intensity slice in the $(0,k,l)$ plane.  Magnetic reflections are present at $k=\pm 0.5$ with
  $l=\pm 1$, $\pm 2$ and $\pm 4$.  The line-cut across $k=0.5$ shows a distinct
  distribution of intensities corresponding to the two structural domains.
  }
  \label{magnetic-peaks}
\end{figure}

\begin{figure}[!ht]
   \includegraphics[width=3.4in]{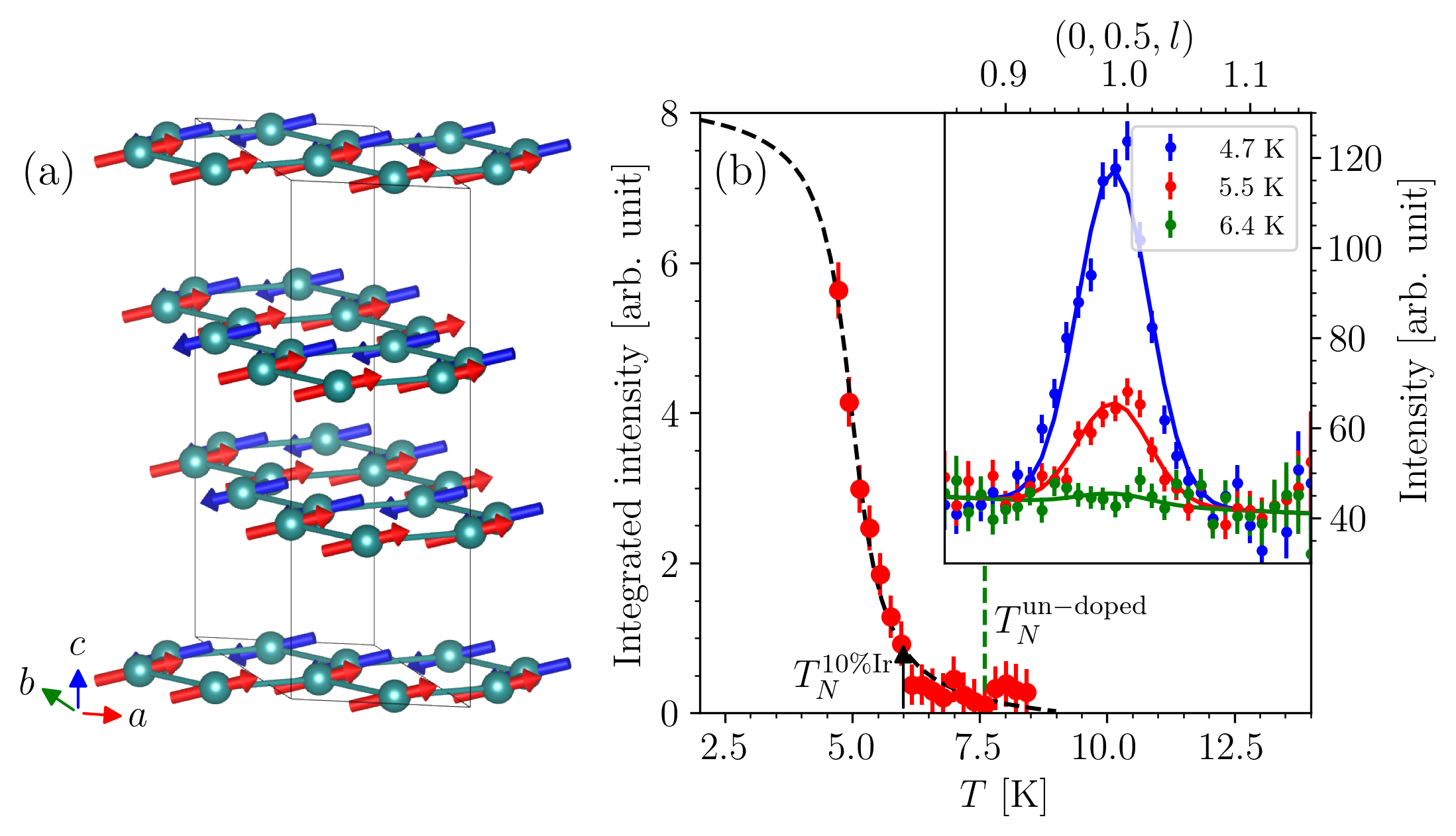}
   \caption{(a) Refined zigzag spin structure of 10\% Ir-substituted $\alpha$-RuCl$_3$ at $T=5.0$ K.
   The magnetic moments are aligned in the $ac$ plane with a canting angle of 15(4)$^\circ$ away
   from the basal plane. (b) The temperature dependence of the integrated intensity of the
   $(0, 0.5, 1)$ magnetic reflection. The dashed line is guide to eye. The inset
   shows selective cuts across the magnetic transition.}
   \label{order-parameter}
\end{figure}

Figures~\ref{intensity-maps}(c)-\ref{intensity-maps}(h) show the reciprocal space contour
maps for both crystals at 200~K (pristine), 5~K (cooled), and 200~K (warmed) conditions.
The $C2/m \rightarrow R\bar{3}$ structural
transition induces notable structural diffuse scattering in the low-temperature phase. The line-cuts
across a Bragg peak along the $c^\ast$-direction in
Figs.~\ref{intensity-maps}(d) and \ref{intensity-maps}(g) for Rh-doping
and Ir-doping, respectively, show clear Lorentzian profile
with a full width half maximum approximately 0.1 reciprocal lattice unit indicating a correlation length of
155$\pm5~\rm\AA$. However, the samples fully recover the original states after
thermal cycling evidenced by the sharp Bragg reflections and
the lack of rod-like feature along the $c^*$-direction.
For the crystal with Rh-doping, this is shown in Figs.~\ref{intensity-maps}(c) and \ref{intensity-maps}(e)
that can be compared to the crystal with Ir-doping shown Figs.~\ref{intensity-maps}(f) and \ref{intensity-maps}(h).
Two distinct sets of Bragg intensities are in the low temperature
trigonal phase: one strong and the other weak. The strong set is from the major structural domain
which fulfills the rhombohedral obverse setting ($-h+k+l=3n$).
The weaker set originates from a twinned domain generated by a two-axis rotation
about the $[1,\bar{1},0]$-direction of the major domain in the reverse setting ($h-k+l=3n$).
This type of obverse/reverse twinning is found to occur in the $R\bar{3}$ space group \cite{herbst02}.
The population ratio is 77:23.
A proper identification of this ratio is important for determining the spin configuration
that relies on an accurate account of all structural and magnetic domain contributions.

Both 10\% substituted compositions enter a magnetic ordered state at low temperature. The characteristic
propagation vector $(0,0.5,1)$ is the same as that reported in the parent
$\alpha$-RuCl$_3$ \cite{park16,balz21} but different from the sample that retains a
low-$T$ monoclinic structure \cite{johnson15}. A complete reciprocal space map is necessary
to elucidate the coexisting structural and magnetic domains for a reliable determination of the
magnetic structure. Figure~\ref{magnetic-peaks}(a) shows the $(h,k,l=1)$ slice of the 10\% Ir-substituted sample at 5~K.
For the major structural domain, there could exist three magnetic twins that are 120$^\circ$
apart along the $c^*$-axis, since the magnetic wavevector breaks the three-fold symmetry.
For this domain, two strong magnetic reflections located at $(0, 0.5, 1)$
and $(0.5, -0.5, 1)$ are visible and highlighted in closed circles.
A third magnetic domain originating from the major
structural domain is hardly visible at the expected $(-0.5,0,1)$ peak position.
By contrast, the minor structural domain shows one distinct magnetic peak at $(0,-0.5,1)$.
Figure~\ref{magnetic-peaks}(b) shows the $(h=0,k,l)$ slice where the magnetic satellite peaks are present
at $k=\pm0.5$. The line-cut along the $[0,0,l]$-direction shows markedly strong intensities
at $l=-2,1$ and $4$, while the weak reflections at remaining $l$ values come from
the minor structural domain. Two structural domains are also observed in the 10\% Rh-substituted sample at 5~K.
The magnetic reflections can also be indexed with
the same $(0, 0.5, 1)$ wavevector, but the magnetic signal is too weak
due to lower counting statistics for proper extraction of the magnetic intensities needed for structure refinement.
More details related to 10\% Rh-substituted sample are given in the Appendix.

The magnetic space group compliant with the wavevector $(0,0.5,1)$
and Ru-ion at the $6c$ site in the parent $R\bar{3}$ space group is $P_S\rm{-}1$ (\#2.7 BNS setting).
Although the magnetic cell doubles the unit cell along the $b$-axis,
there is only one magnetic site with $a$- and $c$-axis components to be determined for the zigzag spin structure.
Given that magnetic intensity measurements probe the spin component perpendicular ${\bf S}_{\perp}$
to the momentum transfer ${\bf Q}$ [${\bf S}_{\perp}=\hat{\bf Q}\times ({\bf S} \times \hat{\bf Q})$],
a comprehensive examination of
out-of-plane $l$ components in the surveyed magnetic reflections $(0,0.5,l=1+3n)$
further constrains the spin canting angles.
A sizable intensity at $l=4$ implies the spin
component is directed closer to the $ab$-plane.
Together with magnetic reflections from other twinned magnetic domains,
magnetic structure refinement of the Ir-substituted
sample using Fullprof \cite{rodriguez-carvajal93} results in a
zigzag structure with spin canting angle slightly away from the basal plane.
The goodness of fit remains relatively flat for the canting angle between -15$^\circ$
to 15$^\circ$  but increases sharply beyond these values. This sets an upper boundary for the canting angle. A more detailed analysis is given in the Appendix.
The ordered moment is determined to be 0.32(5) $\mu_B/\rm Ru$ at 5~K and the resulting
spin configuration is illustrated in Fig.~\ref{order-parameter}(a).
The thermal evolution of the integrated $(0, 0.5, 1)$ peak intensity
is shown in Fig.~\ref{order-parameter}(b) where the N\'{e}el temperature is observed
near 6~K and is consistent with the magnetic susceptibility study in Fig.~\ref{Mag}(a). An earlier study of pure
$\alpha$-RuCl$_3$ \cite{park16} reported an ordered moment of 0.73~$\rm \mu_B/Ru$ at 4~K
with canting angle of 48$^\circ$ from the $a$-axis. The moment at 5~K from this report
based on the change in magnetic Bragg peak is estimated to be 0.69~$\rm \mu_B/Ru$.
In a separate independent study, the moment size at 4.2~K is determined to be 0.45(5)~$\rm \mu_B/Ru$ with canting
angle 35$^\circ$ away from the honeycomb plane \cite{cao16}.  This indicates the magnetic moment size
of the Ir-substituted $\alpha$-RuCl$_3$ sample observed in this study is smaller than that reported in the parent compound $\alpha$-RuCl$_3$.

The spin direction is crucial to understand the strength of relative interactions
and nature of the zigzag order. Since the discovery of $\alpha$-RuCl$_3$ as one of the promising candidates
realizing Kitaev quantum spin liquid, extensive efforts have focused on identifying a proper microscopic
model that is capable of capturing the essential features of the inelastic neutron scattering and
bulk property data \cite{ran17,winter17,wu18,cookmeyer18,lampen-kelley18,ozel19,sears20}.
A combined analysis of various experimental observations has converged on
a minimal generalized Kitaev-Heisenberg Hamiltonian ($K$-$J$-$\Gamma$-$\Gamma^\prime$-$J_3$)
encompassing Ising-like Kitaev interaction $K$, Heisenberg exchange $J$, symmetric off-diagonal
exchange coupling $\Gamma$, $\Gamma^\prime$, and the third-neighbor
Heisenberg coupling $J_3$ \cite{chaloupka16,kim16a,winter16a,hou17,wang17,eichstaedt19,maksimov20,laurell20}.
The Kitaev term $K$ for $\alpha$-RuCl$_3$ is generally considered to be ferromagnetic \cite{winter16a,banerjee16,kim16a,wu18a}
and it is believed the zigzag order is stabilized by either longer-range antiferromagnetic
couplings $J_3$ or anisotropic off-diagonal $\Gamma$ and $\Gamma^\prime$ terms \cite{chaloupka16}.
A sufficiently strong anisotropic term $\Gamma$ leads to a moment locked in the $ac$-crystallographic plane
while the introduction of $\Gamma'$, usually small and originating from trigonal compression,
lowers the critical threshold of $\Gamma$ required to stabilize the moment.
The canting angle $\alpha$ away from the honeycomb plane can be analytically expressed
as $\tan 2\alpha=4/\sqrt{2}(1+r)/(7r-2)$ where $r=-\Gamma/(K+\Gamma')$ \cite{chaloupka13,chaloupka16}.
The canting angle obtained in this study serves as a valuable constraint for understanding the
relative strength of the terms $K$, $\Gamma$, and $ \Gamma'$ in further studies of the
microscopic models of substituted $\alpha$-RuCl$_3$.

Furthermore, our structure investigation reveals minimal alteration of the lattice parameters and
average local RuCl$_6$ environment when compared to the parent compound $\alpha$-RuCl$_3$ (see Appendix).
The trigonal crystal electric field arising from compression of the RuCl$_6$ octahedra, plays
a pivotal role in stabilizing the zigzag order evidenced from
the strong anisotropic magnetization \cite{kubota15},
akin to observations in iridates \cite{gretarsson13a}.
The absence of significant changes in these structural parameters
suggests the suppression of both the transition temperature and ordered moment
comes factors other than chemical pressure \cite{bastien22} or alterations of crystalline
electric field. One possible contributing factor could be the disruption
of exchange pathways due to the dilution of the honeycomb lattice.
On the other hand, the inelastic neutron scattering spectra have shown
that the spin liquid state in $\rm Ru_{1-x}Ir_xCl_3$ is robust against iridium doping;
the low-$T$ characteristic upper excitation feature associated with the fractionalized
excitations is present in a wide range of nonmagnetic iridium doping even when the
static spin order is completely absent \cite{lampen-kelley17}. This is in line with the
recent theoretical study where the introduction of nonmagnetic vacancy concentrations
preserves most of the spin-liquid behavior \cite{kao21}. The vacancies contribute to a pileup
of Majorana modes accumulated in the density of states, which can be experimentally verified
by future low-$T$ specific heat measurements.

\section{Summary}
In summary, we use neutron single crystal diffraction to determine the crystal and magnetic structure
of honeycomb lattice $\alpha$-RuCl$_3$ with Ru partially substituted by nonmagnetic Rh$^{3+}$ and Ir$^{3+}$ ions.
Similar to $\alpha$-RuCl$_3$, both 10\% substituted samples exhibit a zigzag magnetic order with an identical
characteristic magnetic wavevector below $T_N$,  undergo a first-order
structural transition from room-temperature $C2/m$ to low-temperature $R\bar{3}$
phase, fully revert to their original
$C2/m$ state after undergoing thermal cycling. An examination of the nuclear structure
suggests that the substitution of Ir/Rh does not introduce any significant alterations in
either the average structure or the local environment of the RuCl$_6$ octahedra. However,
both the magnetic order and the structural transition upon warming are observed to occur at
lower temperatures in these chemically substituted samples, indicating the nonmagnetic impurities affect both the in-plane magnetic interactions and the interlayer coupling. The magnetic structure refinement
conducted in the Ir-substituted sample has revealed a reduction in both the magnetic moment size
and canting angle when compared to the values reported earlier for $\alpha$-RuCl$_3$.
These findings provide valuable constraints for future theoretical investigations into the
magnetic Hamiltonian in substituted $\alpha$-RuCl$_3$. Chemical substitution with
nonmagnetic ions of similar atomic radii presents an attractive approach for tailoring the
magnetic properties of materials in close proximity to a quantum spin liquid state.

\section{acknowledgments}
IY was supported by an appointment to the Oak Ridge National Laboratory (ORNL)
Next Generation STEM Internship Program (NGSI) Program,
sponsored by the U.S. Department of Energy and administered by the Oak Ridge Institute for Science and Education.
This research was supported by the U.S. Department of Energy (DOE), Office of Science, National Quantum Information Science Research
Centers, Quantum Science Center and used resources at the Spallation Neutron Source, a DOE Office of Science User Facility operated by ORNL. All drawings of crystal and magnetic structures were obtained using VESTA software \cite{momma11}. This manuscript has been authored by UT-Battelle, LLC, under Contract No. DE-AC0500OR22725 with the U.S. Department of Energy.

\section{appendix}
\subsection{Structural Analysis}

Nuclear Bragg peaks of the sample are integrated in reciprocal space and corrected with Lorentz and
spectrum corrections \cite{michels16}. Using a best-fitting three dimensional ellipsoid envelope, the resulting
intensities are corrected for wavelength-dependent absorption \cite{sears92} of the sample using an equivalent sphere
analytical correction \cite{dwiggins75} with chemical formula $\rm Ru_{0.9}{\it R}_{0.1}Cl_3$ ($\rm {\it R}=Ir, Rh$),
unit cell $Z$-parameter and volume.
The Mantid software is used throughout the data reduction process \cite{arnold14}.
In the Jana crystal structure refinements, the twin law corresponding to obverse/reverse twinning is introduced
with variable domain population. A wavelength-dependent secondary spherical extinction correction model is applied \cite{becker74}.
Structure refinement of the calculated structure factors to the corrected Bragg intensities
from the complete reciprocal space maps are used to refine the positions of the Ru and Cl-sites.
Determination of the exact doping level is difficult due to the typical broadening of Bragg peaks
observed in van der Waals materials. In addition, the scattering length of Rh is somewhat closer to Ru
than Ir. For the 10\%-samples measured at CORELLI, the compositions are determined to be 6-7\% for Ir-doped sample
and 11\% for Rh-doped sample. Subsequent refinements to obtain the bond distances and angles
are done with composition fixed at $x=0.1$.
For the TOPAZ Rh-doped sample with the nominal composition of $x=0.2$, the sample composition is determined to be 22\% Rh..

\begin{table}[!ht]
\caption{Structural parameters at high and low temperature phases
for the doped $\alpha$-RuCl$_3$ compared to un-doped one. Note uncertainties of
lattice constant are standard errors. Lengths are in angstroms and angles are in degrees}
\centering
\begin{ruledtabular}
\begin{tabular}{lllll}
$C2/m$ & 10\%-Rh & 20\%-Rh  & 10\%-Ir & parent \cite{zhang23} \\
\hline
$a$ & 6.000(5) & 5.980(1) & 5.981(3) & 5.976(1) \\
$b$ & 10.331(4) & 10.348(1) & 10.344(2) & 10.343(1) \\
$c$ & 6.0092(2) & 6.0356(1) & 6.027(1) & 6.023(1) \\
$\beta$ & 108.74(5) & 108.72(1) &  108.88(2) & 108.80(1) \\
\hline
$R\bar{3}$ & 10\%-Rh & 20\%-Rh  & 10\%-Ir& parent \cite{park16}\\
\hline
$a$ & 5.973(3) & 5.974(1) & 5.972(1) & 5.973(1) \\
$c$ &  16.956(9) & 16.961(2) & 16.953(3) &  16.93(6)\\
\end{tabular}
\end{ruledtabular}
\label{app:lattice-constants}
\end{table}

\begin{table}[!h]
\caption{Selected Ru-Cl bond distances and angles at high and low temperature phases
for the doped $\alpha$-RuCl$_3$ compared to un-doped one. Note uncertainties are
standard errors. Lengths are in angstroms and angles are in degrees.}
\centering
\begin{ruledtabular}
\begin{tabular}{llllll}
$C2/m$ & 10\%-Rh & 20\%-Rh  & 10\%-Ir & parent \cite{zhang23} \\
\hline
Ru1-Cl1 & 2.360(2) & 2.363(1) & 2.361(2) & 2.363(1) \\
Ru1-Cl2 & 2.358(2) & 2.361(1) & 2.361(2) & 2.361(1)\\
Ru1-Cl2$^{\rm ii}$ & 2.359(2) & 2.359(1) & 2.359(2) & 2.359(1) \\
\hline
$R\bar{3}$ & 10\%-Rh & 20\%-Rh  & 10\%-Ir& parent \cite{park16}\\
\hline
Ru1-Cl1 & 2.363(3) & 2.361(4) & 2.362(1) & 2.359(6) \\
Ru1-Cl1$^{\rm III}$ & 2.356(2) & 2.360(3) & 2.359(1) & 2.352(6) \\
\hline
$C2/m$ & 10\%-Rh & 20\%-Rh  & 10\%-Ir & parent \cite{zhang23} \\
\hline
Cl1-Ru1-Cl1$^{\rm i}$ & 86.50(7) & 86.27(3) & 86.37(5) & 86.48(2) \\
Cl1-Ru1-Cl2 & 91.30(7) & 91.61(3) & 91.48(6) & 91.55(2) \\
Cl1-Ru1-Cl2$^{\rm i}$ & 91.49(7) & 91.10(3) & 91.25(6) & 91.20(2) \\
Cl1-Ru1-Cl2$^{\rm ii}$ & 91.14(6) & 91.30(2) & 91.33(4) & 91.24(2) \\
Cl2-Ru1-Cl2$^{\rm ii}$ & 85.58(6) & 85.99(2) & 85.92(4) & 86.01(2) \\
Cl2-Ru1-Cl2$^{\rm iii}$ & 91.75(6) & 91.40(2) & 91.46(4) & 91.35(2) \\
Cl2$^{\rm ii}$-Ru1-Cl2$^{\rm iii}$ & 91.36(7) & 91.28(3) & 91.11(5) & 91.18(2) \\
\hline
$R\bar{3}$ & 10\%-Rh & 20\%-Rh  & 10\%-Ir& parent \cite{park16}\\
\hline
Cl1-Ru1-Cl1$^{\rm I}$ & 91.39(6) & 91.56(8) & 91.58(2) & 91.36(2) \\
Cl1-Ru1-Cl1$^{\rm III}$ & 86.09(5) & 86.12(8) & 86.15(2) & 85.90(2) \\
Cl1-Ru1-Cl1$^{\rm V}$ & 91.24(5) & 91.24(8) & 91.17(2) & 91.61(1) \\
Cl1$^{\rm III}$-Ru1-Cl1$^{\rm IV}$ & 91.39(6) & 91.18(8) & 91.21(2) & 91.26(2) \\
\end{tabular}
\end{ruledtabular}
\label{bonds}
\raggedright
\footnotesize{Superscript indicates symmetry operator. Unlabeled refers
to $(x,y,z)$. $C2/m$: $^{\rm i}(-x,y,-z+1)$, $^{\rm ii}(-x+\sfrac{1}{2},-y+\sfrac{1}{2},-z+1)$,
$^{\rm iii}(x-\sfrac{1}{2},-y+\sfrac{1}{2})$; $R\bar{3}$: $^{\rm I}(-y,x-y,z)$,$^{\rm II}(-x+y,-x,z)$,
$^{\rm III}(-x+\sfrac{1}{3},-y+\sfrac{2}{3},-z+\sfrac{2}{3})$,
$^{\rm IV}(y-\sfrac{2}{3},-x+y-\sfrac{1}{3},-z+\sfrac{2}{3})$,
$^{\rm V}(x-y+\sfrac{1}{3},x-\sfrac{1}{3},-z+\sfrac{2}{3})$}
\end{table}

Table~\ref{app:lattice-constants} lists the lattice parameters of both the high temperature $C2/m$  and low temperature $R\bar{3}$ phases. The parameters for all compositions appear similar. For the monoclinic
case, a reference sample previously measured on CORELLI at 200~K is reported \cite{zhang23}.
In the case of the trigonal phase, the parameters at 5~K from \cite{park16} are used for comparison.
For clarity, the $R\bar{3}$ space group has a hexagonal unit cell with $R$ rhombohedral lattice centering.
Structural refinement analysis is performed using Jana2020 \cite{petricek23} to obtain the
refined site fractional coordinates. Typical bond-distances, and angles
of the RuCl$_6$ octahedra are reported in Table~\ref{bonds}. In this table, the superscript
indicates a symmetry operator of the site that forms the bond. No superscript
indicates the identity operator $(x,y,z)$. There does not appear to be a significant
difference among all samples studied. The low temperature
reference values for the trigonal case are also from \cite{park16} at 5~K.

\subsection{Magnetic Structure Analysis}

\begin{table}[!ht]
\caption{Basis vectors of irreducible representation of zigzag magnetic structure with the Ru $6c$ site.}
\begin{ruledtabular}
\begin{tabular}{lccc}
$\Gamma_1$ & $\psi_1$ & $\psi_2$ & $\psi_3$\\
\hline
Ru $(0,0,z)$ & $(1,0,0)$ & $(0,1,0)$ & $(0,0,1)$ \\
Ru $(0,0,\bar{z})$ & $(\bar{1},0,0)$ & $(0,\bar{1},0)$ & $(0,0,\bar{1})$
\label{basis}
\end{tabular}
\end{ruledtabular}
\end{table}

\begin{figure}[!ht]
   \includegraphics[width=3.4in]{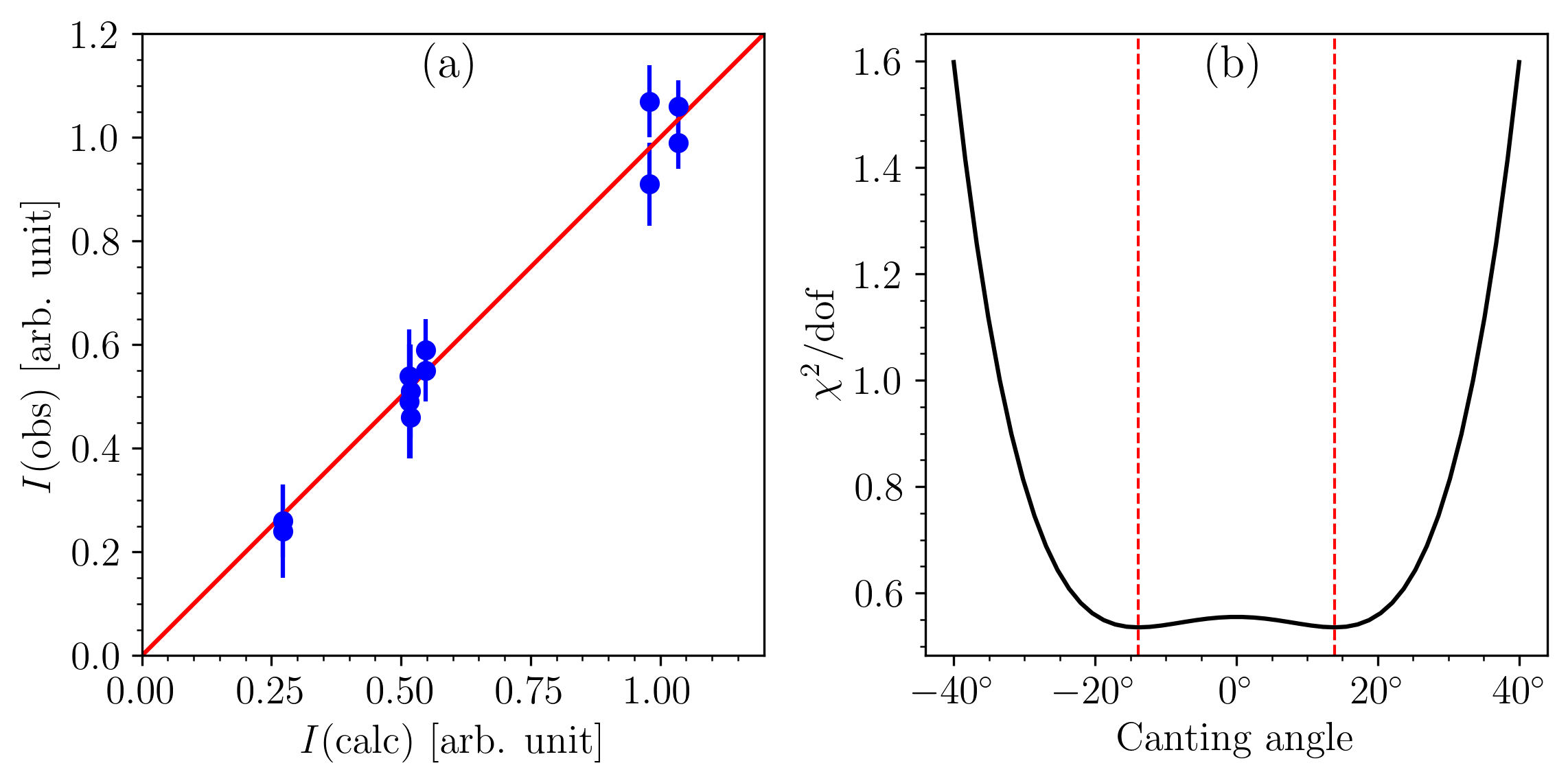}
   \caption{(a) The observed to calculated intensities with canting
   angle that minimizes the $\chi^2$ from 12 magnetic peaks. The solid red line is guide to the eye.
   (b) Dependence of fitted $\chi^2$ per degree of freedom with canting angle comparing observed
   integrated intensities to calculated structure factors for the 10\% Ir-substituted sample. Canting angle
   shows a clear minimum near 15$^\circ$ denoted with the dashed red line.}
   \label{F2-calc}
\end{figure}

The zigzag magnetic structure corresponds to irreducible representation $\Gamma_1$.
The basis vectors from representation analysis are shown in Table~\ref{basis} for the Ru ion and used in the
FullProf refinement \cite{rodriguez-carvajal93}. Magnetic peaks from the Ir-substituted sample are integrated in a similar
fashion to the structural Bragg peaks. Six magnetic peaks from each structural domain are integrated.
The primary domain with 77\% contribution contains the
magnetic reflections that are offset from indexed nuclear peaks in the rhombohedrally centered obverse setting.
The secondary domain with 23\% has peaks indexed from the reverse twin as discussed in the text.
To determine the moment size, the magnetic peaks are refined with the Bragg reflections
to obtain the scale factor.

The observed magnetic intensities are compared to the expected values for the
refined structure are shown in Fig.~\ref{F2-calc}(a) for the Ir-substituted sample with canting angle 15$^\circ$.
The $\chi^2$ goodness of fit per degree of freedom (dof) with variable
canting angle from the magnetic structural model is
shown in Fig.~\ref{F2-calc}(b) and shows a clear minimum near 15$^\circ$.
Above $15^\circ$ (and below $-15^\circ$), the $\chi^2$ deviates drastically.
Hence, $15^\circ$ is the upper bound of the canting angle.
As the change in $\chi^2$ is small about zero canting angle within $\pm15$
does not rule out smaller canting angles within uncertainty.

\begin{figure}[!ht]
   \includegraphics[width=3.4in]{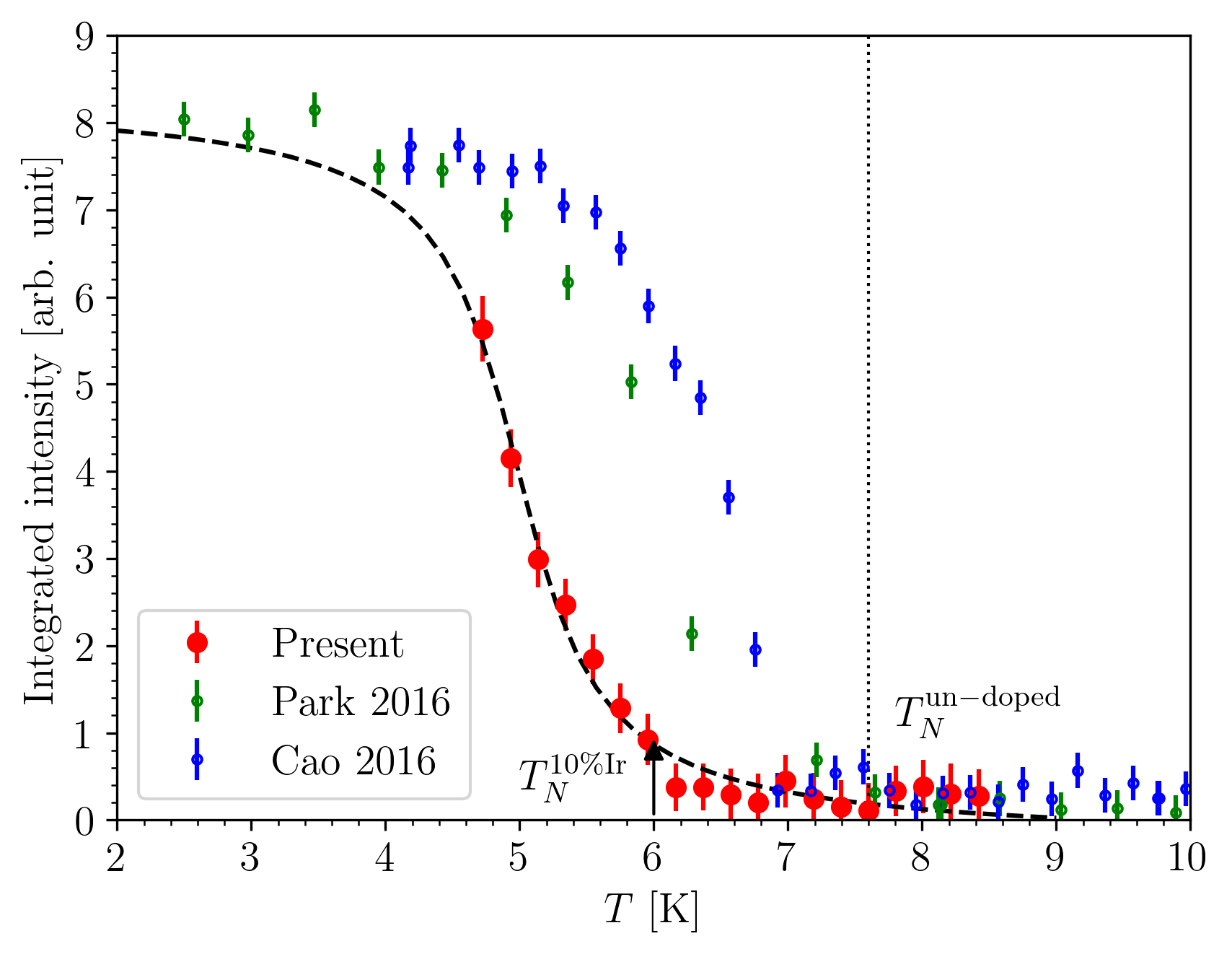}
   \caption{Comparison of the thermal evolution of the 10\% Ir-substituted $(0,0.5,1)$ magnetic satellite peak
   to order parameters from references \cite{park16,cao16} corresponding to un-substituted $\alpha$-RuCl$_3$.
   The intensity values of the reference data are scaled to the guide line.
   The transition of the present data appears lower than the parent material.
   }
   \label{order-parameter-comparison}
\end{figure}

The order parameter of the Ir-substituted $(0,0.5,1)$ magnetic satellite peak
integrated intensity is compared to reference data for the parent compound \cite{park16,cao16} in
Fig.~\ref{order-parameter-comparison}. The intensities of the reference data
are scaled by a factor to the guide line. For reference \cite{park16} with un-substituted
crystal structure determined in space group $R\bar{3}$, it is estimated that
the moment size of the parent compound at 5~K data will be about 0.69~$\rm \mu_B/Ru$ which
is associated with an approximately 10 \% drop in intensity from the value at 4~K estimated
from the intensity reduction of the order parameter.
A reduction of intensity by a factor is associated with a corresponding change of
moment size by the square root of that factor since intensities scale with the square of the magnetic
structure factor.
Similarly, for reference \cite{cao16} with magnetic structure determined in parent space group $C2/m$,
the associated drop in intensity is approximately 5\% at 5K indicating a minor change in moment size
(approximately 0.46~$\rm \mu_B/Ru$). This suggests the magnetic moment size of the Ir-substituted sample
is smaller than these previous reports.

\begin{figure}[h!t]
   \includegraphics[width=3.4in]{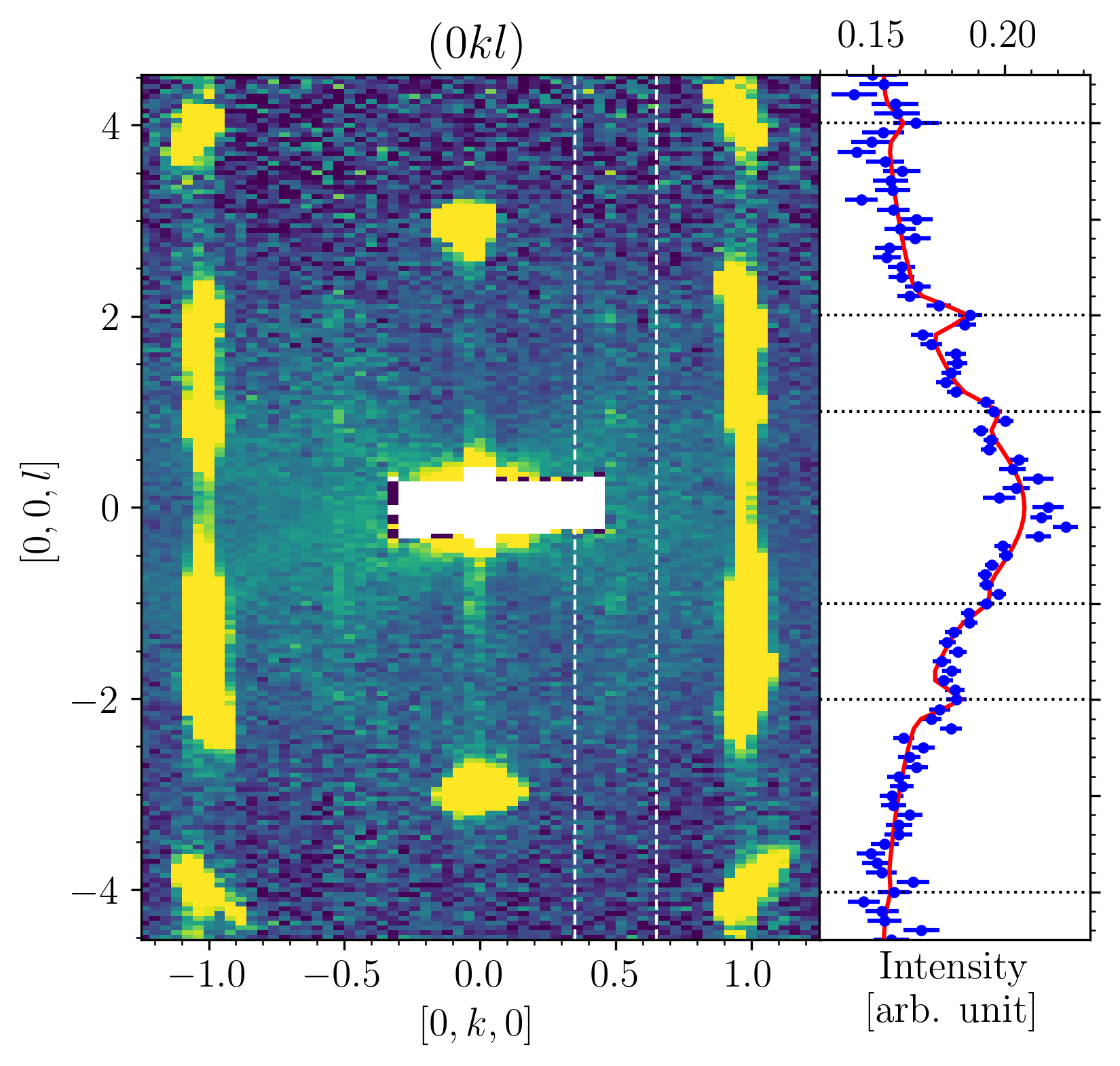}
   \caption{The intensity slice in the $(0,k,l)$ plane from the 10\% Rh-substituted sample. Magnetic reflections are
   present but appear weaker than observed in 10\% Ir-doping at $k=\pm 0.5$ with $l=\pm 1$, $\pm 2$ and $\pm 4$.
   }
   \label{rh-intensity}
\end{figure}

As the Rh-doped sample has a similar reverse/obverse structural twinning with one major and minor domain,
the appearance of magnetic peaks follows a similar trend with the Ir-doped sample.
The $(0,k,l)$ intensity slice from the Rh-doped sample at 5~K is
displayed in Fig.~\ref{rh-intensity} and shows magnetic peaks appearing at similar positions
as the Ir-doped sample out to $l=\pm4$ which include
$(0,0.5,\pm1)$, $(0,0.5,\pm2)$, and $(0,0.5,\pm4)$. Magnetic peaks can also be observed along
the $[0,-0.5,l]$ direction. Due to counting statistics,
the magnetic peaks are too weak for reliable integration and magnetic structure refinement.

%\bibliography{alpha}

%\widetext
%\clearpage

%\appendix

%\input{supporting.tex}

\end{document}